\documentclass{aa}  

\usepackage{graphicx}
\usepackage{txfonts}
\usepackage{hyperref}
\usepackage{newtxtext,newtxmath}

\usepackage[T1]{fontenc}
\usepackage[normalem]{ulem}
\usepackage{color}

\DeclareRobustCommand{\VAN}[3]{#2}
\let\VANthebibliography\thebibliography
\def\thebibliography{\DeclareRobustCommand{\VAN}[3]{##3}\VANthebibliography}

\usepackage{amsmath}	
\usepackage{cuted}

\newcommand{\sdt}[1]{{\color[rgb]{0.9,0.0,0.0} {\textbf{#1}}}}

\begin{document}

   \title{ZTF SN~Ia DR2: Cosmology-independent constraints on Type Ia supernova standardisation from supernova siblings}
\titlerunning{ZTF Siblings SN~Ia}

   \author{S. Dhawan,
          \inst{1}
          E. Mortsell\inst{2},
          J. Johansson\inst{2} A. Goobar,\inst{2} M. Rigault,\inst{3},  M. Smith \inst{3,4}, K. Maguire,\inst{5} J. Nordin,\inst{6} G. Dimitriadis\inst{5}, P.E. Nugent\inst{7,8}, L. Galbany,\inst{9,10} J. Sollerman,\inst{11} 
T. de Jaeger,\inst{12}, J.H. Terwel,\inst{5,13} Y.-L. Kim \inst{4}, Umut Burgaz \inst{5}, G. Helou \inst{14} , J. Purdum \inst{14}, S. L. Groom \inst{14}, R. Laher \inst{14} \and B. Healy \inst{15,16}
          }
   \institute{Institute of Astronomy and Kavli Institute for Cosmology, University of Cambridge, Madingley Road, Cambridge CB3 0HA, UK\\
              \email{suhail.dhawan@ast.cam.ac.uk}
         \and
           The Oskar Klein Centre for Cosmoparticle Physics, Department of Physics, Stockholm University, SE-10691 Stockholm, Sweden
           \and 
           Université de Lyon, Université Claude Bernard Lyon 1, CNRS/IN2P3, IP2I Lyon, F-69622, Villeurbanne, France 
           \and 
            Department of Physics, Lancaster University, Lancs LA1 4YB, UK 
            \and 
           School of Physics, Trinity College Dublin, The University of Dublin, Dublin 2, Ireland 
           \and
           Institut für Physik, Humboldt-Universität zu Berlin, Newtonstr. 15, 12489 Berlin, Germany
           \and
          Lawrence Berkeley National Laboratory, 1 Cyclotron Road MS 50B-4206, Berkeley, CA, 94720, USA
           \and
         Department of Astronomy, University of California, Berkeley, 501 Campbell Hall, Berkeley, CA 94720, USA
           \and 
           Institute of Space Sciences (ICE, CSIC), Campus UAB, Carrer de Can Magrans, s/n, E-08193 Barcelona, Spain 
           \and 
           Institut d’Estudis Espacials de Catalunya (IEEC), E-08034 Barcelona, Spain 
           \and
           The Oskar Klein Centre, Department of Astronomy, Stockholm University, Albanova University Center, Stockholm, SE-106 91, Sweden 
           \and 
           LPNHE, (CNRS/IN2P3, Sorbonne Universit\'{e}, Universit\'{e} Paris Cit\'{e}), Laboratoire de Physique Nucl\'{e}aire et de Hautes \'{E}nergies, 75005, Paris, France
           \and 
           Nordic Optical Telescope, Rambla José Ana Fernández Pérez 7, ES-38711 Breña Baja, Spain 
           \and 
           IPAC, California Institute of Technology, Pasadena, CA 91125, USA 
             \and 
             Caltech Optical Observatories, California Institute of Technology, Pasadena, CA 91125, USA 
             \and
             School of Physics and Astronomy, University of Minnesota, Minneapolis, MN 55455, USA} 
\authorrunning{S. Dhawan et al. }
 
  \abstract{
Understanding Type Ia supernovae (SNe~Ia) and the empirical standardisation relations that make them excellent distance indicators is vital to improving cosmological constraints. SN~Ia ``siblings", i.e. two or more SNe~Ia in the same host or parent galaxy offer a unique way to infer the standardisation relations and their diversity across the population. We analyse a sample of 25 SN~Ia pairs, observed homogeneously by the Zwicky Transient Factory (ZTF) to infer the SNe~Ia light curve width-luminosity and colour-luminosity parameters $\alpha$ and $\beta$. Using the pairwise constraints from siblings, allowing for a diversity in the standardisation relations, we find $\alpha =  0.218  \pm  0.055 $ and $\beta = 3.084  \pm  0.312$, respectively, with a dispersion in $\alpha$ and $\beta$ of $\leq  0.195$ and $\leq 0.923$, respectively, at 95$\%$ C.L. While the median dispersion is large, the values  within $\sim 1 \sigma$  are consistent with no dispersion. Hence, fitting for a single global standardisation relation, we find $\alpha = 0.228  \pm  0.029 $ and $\beta =  3.160  \pm  0.191$. We find a very small intrinsic scatter of the siblings sample $\sigma_{\rm int} \leq 0.10$ at 95\% C.L. compared to  $\sigma_{\rm int} = 0.22 \pm 0.04$ when computing the scatter using the Hubble residuals without comparing them as siblings.  Splitting the sample based on host galaxy stellar mass, we find that SNe~Ia in both subsamples have consistent $\alpha$ and $\beta$. The $\beta$ value is consistent with the value for the cosmological sample. However,  we find a higher $\alpha$ by $\sim 2.5 - 3.5 \sigma$. The high $\alpha$ is driven by low $x_1$ pairs, potentially suggesting that the slow and fast declining SN~Ia have different slopes of the width-luminosity relation. We can confirm or refute this with increased statistics from near future time-domain surveys. If confirmed, this can both improve the cosmological inference from SNe~Ia and infer properties of the progenitors for subpopulations of SNe~Ia.}

   \maketitle
%

\section{Introduction}
Type Ia supernovae (SNe~Ia) are excellent distance indicators in cosmology, instrumental in the discovery of the accelerated expansion of the universe \citep{Riess:1998cb,Perlmutter:1998np}. SNe~Ia are crucial to measuring dark energy and the Hubble constant, precisely \citep[e.g.][]{Brout2022,Riess2022}. In optical wavelengths, the regime where most constraints on cosmology from SNe~Ia are obtained, standardisation of their peak luminosity can reduce the scatter to $\sim 15\%$. The peak brightness is corrected for correlations with the lightcurve width and colour \citep[e.g.,][]{phillips1993,tripp1998} and also host galaxy properties \citep[e.g.,][]{Kelly2010,Sullivan2010}. The dependence of width and colour corrected luminosity on host galaxy  stellar mass, commonly termed the ``mass step" is crucial for improving cosmological constraints. The origin of the mass step has been poorly understood, but recent studies \citep[e.g.][]{BS20} suggest this could be due to dust and / or intrinsic differences related to astrophysical properties, e.g. progenitor age \citep{rigault2020,briday2022}.  As SN~Ia cosmology is currently systematics limited, understanding the standardisation relations is crucial to constrain cosmology. This is particularly important since several future Stage-IV dark energy missions are designed with a sizable component devoted to a high-redshift SN~Ia survey \citep[][]{Hounsell2018,SRD_LSST}. At low-redshift a large sample of well-characterised SNe~Ia has already been obtained by surveys like the Zwicky Transient Facility \citep[ZTF;][]{graham2019,bellm2019,Dekany2020}.

SN~Ia siblings, i.e. multiple SNe~Ia in the \emph{same} parent galaxy \citep{Brown2015}, are a powerful route to constrain these standardisation relations. Recently, SN~Ia cosmological samples have been analysed using the SALT2 model \citep{guy2007,guy2010}, wherein the distance modulus $\mu$ is obtained by correcting the inferred apparent peak magnitude ($m_B$) for the lightcurve width ($x_1$),  and colour ($c$) by the relation 
\begin{equation}
\mu = m_B + \alpha x_1 - \beta c - M_B, 
\label{eq:tripp}
\end{equation}
where $\alpha$ and $\beta$ are derived from a simultaneous fit along with cosmology to minimize scatter in the Hubble-Lemaitre diagram. In the SALT2 formalism, $c$ is an observed colour, which can be viewed as a combination of the intrinsic colour and dust. 

The parameter $\beta$ - central to this work - is empirically derived and captures both the intrinsic and extrinsic colour-luminosity relations. In terms of the latter, $\beta$ can be viewed as an analog of the total-to-selective absorption ratio in the $B$-band, $R_B$, for a given dust law \citep{1989ApJ...345..245C}.
A simultaneous cosmology fit using the largest compiled sample of SNe~Ia, inferred $\beta = 3.04 \pm 0.04$ \citep{Brout2022}, significantly lower than the $R_B \sim 4.1$ seen in the Milky Way  \citep{1989ApJ...345..245C,Fitzpatrick1999}. 
In cosmological surveys of SNe~Ia, it has been noticed that selection effects can lead to incompleteness in the distribution of SNe Ia properties due to correlations with the intrinsic dispersion. These effects will impact the standardisation relations and they are corrected for using simulations \citep[e.g.,][]{Kessler2019,Popovic2021}. These simulations require detailed inputs of survey observations and the population models derived from the data \citep[e.g.][]{Scolnic2016}. 
Exploring the colours of nearby SNe~Ia \citep[e,g,][]{2008A&A...487...19N}, and further expanding the wavelength coverage of the observations from UV to the NIR \citep{2014ApJ...789...32B,2015MNRAS.453.3300A} indicate a wide range of dust distributions in the interstellar medium (ISM) of SN~Ia host galaxies to explain the observed colours.  
The procedure for the cosmological inference of $\alpha$ and $\beta$ is convolved with effects like K-corrections, selection effects, redshift uncertainties, and even Milky Way extinction errors. It is, therefore, important to have independent methods for measuring the standardisation relations. Studies with cosmological samples have shown the likelihood of $\beta$ values to be dependent on the host galaxy environment \citep{Gonz_lez_Gait_n_2021,BS20}, which is crucial for precision inference of cosmology with current and future samples.

Owing to multiple SNe~Ia exploding in the same galaxy, the inference from sibling SNe~Ia is insensitive to certain systematics, e.g. cosmological model parameters, peculiar velocity corrections and global host galaxy dependence. Therefore, it is a robust, independent test of the width-luminosity and colour-luminosity relations, as demonstrated constraining the colour-luminosity relation ($\beta$) from a single sibling pair in \citet{Biswas2022}, where $\beta$ is $3.5 \pm 0.3$. 
In the recent literature, it has been posited that SN~Ia siblings could have a smaller dispersion in their luminosity compared to SNe~Ia in different galaxies \citep{Burns2020}. This is also seen in the small distance dispersion for the three spectroscopically normal SNe~Ia in NGC 1316 \citep{Stritzinger2010}, although the spectroscopically peculiar SN~2006mr  has a distance modulus that differs by 0.6 mag  from that of the other three. Other studies, however, find no difference between the scatter in SN~Ia siblings and non-sibling SNe~Ia \citep{Scolnic2020,Scolnic2022}. Apart from understanding and improving the distance measurements for cosmology, comparing siblings also has interesting implications for SN~Ia physics. \citet{Gall2018} analysed SN2007on and SN2011iv and found an difference of 14$\%$ and 9$\%$ in their distances from the optical and NIR, respectively. This was attributed to the differences in the progenitor systems, hypothesized to be due to different central densities of the primary white dwarf \citep[e.g.][]{Ashall2018}. 
It is, therefore, interesting to study SN~Ia siblings to both understand the luminosity corrections and test whether the absence of potential systematics in common can increase the precision in distance measurements. While we can collect a large sample of historical SN~Ia sibling data \citep[e.g.][]{Anderson2013, Kelsey2023}, studies like \citet{Burns2020} have shown that the systematics from heterogeneous photometric systems add significant dispersion to the distances and the scatter is significantly smaller for a sample observed with the same photometric system. 

In this paper, we analyse a sample of SN~Ia siblings homogeneously observed by ZTF. A large part of the sample of SN~Ia siblings is derived from the second data release of SNe~Ia observed by ZTF (ZTF DR2).  We infer the SALT2 parameters and subsequently the width and colour-luminosity relation as presented in equation~\ref{eq:tripp}. With a sizable sample of siblings, we both present a cosmology independent inference of $\alpha$ and $\beta$ and an estimate of the observed  diversity of both parameters across the sample. Since all SNe~Ia are on the same photometric system, this will minimise cross-calibration systematics, which have been a significant source of error in cosmological studies \citep[e.g.][]{Scolnic2022,Brout2022b}. 
We use the ZTF sample of SN~Ia siblings to infer SN lightcurve parameters and simultaneously constrain the width-luminosity and colour-luminosity relations.   We present the method in section~\ref{sec:method}, the results in section~\ref{sec:results}  and discuss our findings with respect to the literature, specifically, the cosmological sample of SNe~Ia in section~\ref{sec:discussion}. Finally, we conclude in section~\ref{sec:conclusion}.

\section{Data and Methodology}
\label{sec:method}
Initial studies of multiple SNe in the same galaxy often focussed on a single pair or a small set of siblings \citep{Hamuy1991,Stritzinger2010}. With modern time-domain astronomy surveys having a long survey duration, it has been possible to assemble larger samples of SN~Ia siblings \citep{Scolnic2020,Scolnic2022, Burns2020}.
ZTF is an optical imaging survey of the entire Northern sky with a 3-day cadence in the $g$ and $r$ bands with a $\sim$ 20.5 mag depth which operated between 2018 and 2020 and was augmented to a 2-day cadence since 2020 with its successor ZTF-II. This public $g$ + $r$ band survey is complemented with partnership surveys in the $i$ band and higher cadence observations. The unprecedented scanning speed and depth has made ZTF the ideal machinery for discovering and characterising SN siblings \citep{Biswas2022,Graham2022}. Lightcurves for the objects in this paper were built using a variant on the standard IPAC forced photometry pipeline \citep{Masci2019}, with more details in associated papers (e.g. Smith et al. in prep.).

We construct our sibling sample by starting with the sample of spectroscopically confirmed ZTF SNe~Ia. We query the catalog of SNe~Ia for transients, using \texttt{fritz} \citep{Vanderwalt2019,Coughlin2023} within a 100 arcsecond radius from the SN~Ia coordinates (at $z \sim 0.1$ this corresponds to a physical separation of $\sim 35$\, kpc). We then save the sample of pairs for which the second object is also associated to the same host galaxy. From this sample, we remove objects which show continuous variability for more than 60 days before the date of maximum brightness, to remove persistent transient sources like active galactic nuclei (AGNs) and tidal disruption events (TDEs).
Details of which sibling pairs passed the sample selection are presented in section~\ref{sec:sample_select}

\begin{figure}
    \centering
    \includegraphics[width=.5\textwidth]{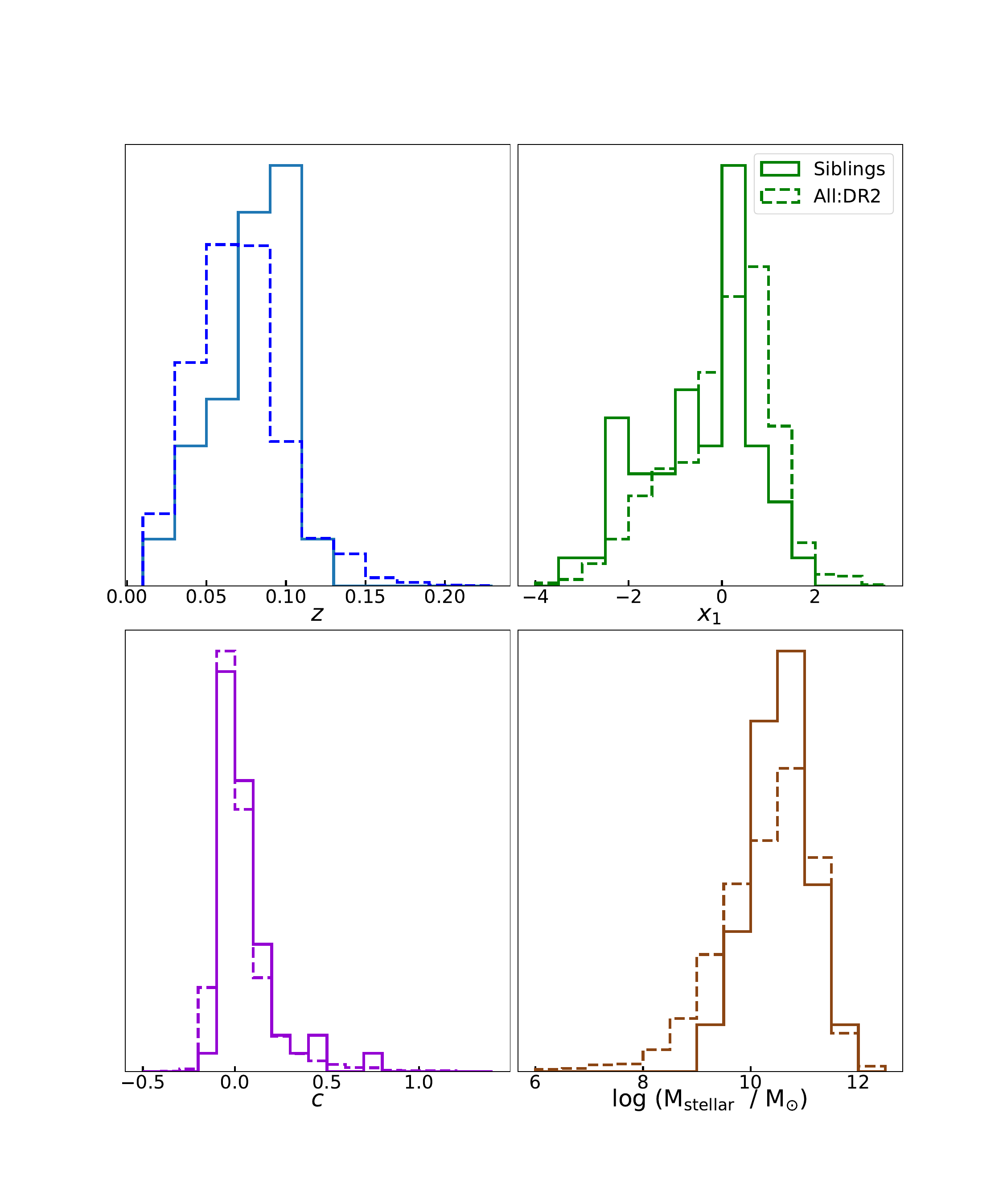}
    \vspace{-1cm}
    \caption{ Parameters for the sibling SNe~Ia in this study. The panels show the redshift (top left), SALT2 $x_1$ (top right), $c$ (bottom left) and host galaxy stellar mass (bottom right) distributions. We do not make any selection cuts on the values of $z$, $x_1$ and $c$, unlike for the cosmological sample. The equivalent distribution for the entire DR2 sample is overplotted as dashed lines. }
    \label{fig:param_dist}
\end{figure}

For our analyses, we take a sample of 24      both spectroscopically classified SN~Ia sibling pairs (hereafter, spec) and 28 pairs with one spectroscopically classified SN~Ia and one photometrically classified SN~Ia (hereafter, photo-spec) that have multi-band lightcurves from ZTF. We describe below the process of determining that the objects in the photo-spec sample without a classification are SNe~Ia. \footnote{This sample includes siblings discovered both in phase I and II of ZTF operations. Henceforth, we refer to both ZTF-I + II as ZTF, for brevity.}
While we do have two subsamples we homogeneously analyse the entire sample with the same assumptions and selection cuts. However, for our analysis, we also make consistency checks between the spec and photo-spec subsamples along with providing the joint constraints. Most of the sibling pairs analysed in this work have at least one member in the second data release (DR2) of ZTF SNe~Ia (\sdt{Rigault et al. in prep.} hereafter R24; \sdt{Smith et al. in prep.}, hereafter S24) and a large fraction of them were classified with the SEDmachine on the Palomar P60 \citep{Blago2018,Rigault2019}. We note that given the approximate rate of one SN~Ia per galaxy per century, the total number of sibling pairs in our sample is consistent given the size of the entire DR2 sample is $\sim 3000$ SNe~Ia.
Since previous studies with SN~Ia siblings \citep[e.g.,][]{Burns2020} suggest that cross-calibration systematics are a large error source in sibling analyses, we only construct our sample from sibling pairs where both SNe are observed by ZTF. Our sample of siblings spans a large redshift range from $0.01 < z < 0.1$, as shown in Figure~\ref{fig:param_dist}. 

\begin{figure}
    \centering
\includegraphics[width=.48\textwidth, trim= 0 20 0 10]{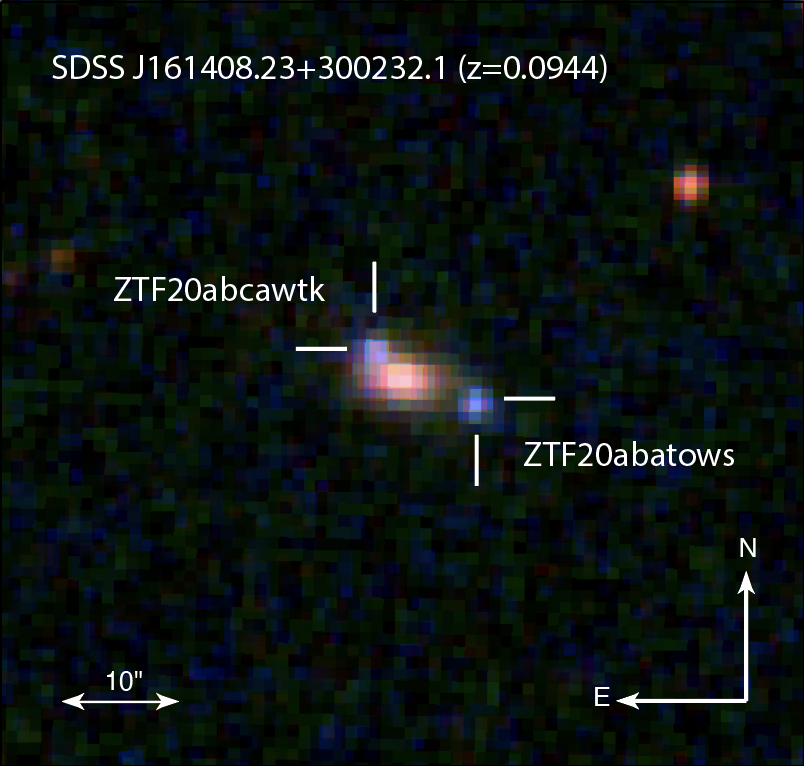}
    \caption{The ZTF RGB image of an example sibling pair from our sample, ZTF20abatows and ZTF20abcawtk. The crosses mark the position of the SNe~Ia in the field. These siblings were closest in the time separation ($\sim 5$ days between peak for the two SNe) between their peaks and hence, were detectable at the same time.}
    \label{fig:image_sibs}
\end{figure}

Currently, the most widely used lightcurve fitting algorithm is the Spectral Adaptive Lightcurve Template - 2 \citep[SALT2;][]{guy2010}, based on the SALT method \citep{2005A&A...443..781G} and we use this in our analysis. The SALT2 model treats the colour entirely empirically and is used to find a global colour-luminosity relation. We use the updated version of SALT2 presented in \citet{Taylor2021} as implemented in \texttt{sncosmo} v2.1.0 \footnote{\url{https://sncosmo.readthedocs.io/en/v2.1.x/}} \citep{2016ascl.soft11017B}. We fit, iteratively, wherein the first iteration without the model covariance is only used to guess the time of maximum. The second iteration is fitting the model to  only the data between -10 and +40 days from the first guess time of maximum (see Rigault et al. in prep. for details on selection of the phase range) and with the model covariance to get all the SALT2 fit parameters simultaneously.  In the fitting procedure, we correct the SN fluxes for extinction due to dust in the Milky Way (MW), using extinction values  derived in \citet{2011ApJ...737..103S}.  We use the widely applied galactic reddening law, proposed in \citet{1989ApJ...345..245C}, known as the ``CCM" law to correct for MW extinction, with the canonical value for the total-to-selective absorption, $R_V = R_B - 1 = 3.1$. We also compare with an older, more widely used version of SALT2 from \citet{Betoule2014} and find consistent estimates of the inferred parameters.  For the SNe in the sample without a spectroscopic classification, we fit template spectral energy distributions for SN~Ib/c, IIn and IIP, as provided very kindly by Peter Nugent\footnote{https://c3.lbl.gov/nugent/nugent templates.html}. Only the pairs where the SN without a spectroscopic classification also prefers a fit to an SN~Ia template (by a $\Delta \chi^2$ of at least 5, though in most cases the fit is overwhelmingly preferring an SN~Ia  by a $\Delta \chi^2$ of $\sim 50$ or greater) are kept in the sample. 

Here, we aim to infer the standardisation relations between the SN~Ia luminosity and the lightcurve width and colour. 
The SALT2 model is typically used with a standardisation relation as parametrised in \citet{tripp1998}

\begin{figure}
    \centering
    \includegraphics[width=.48\textwidth, trim = 0 20 0 10]{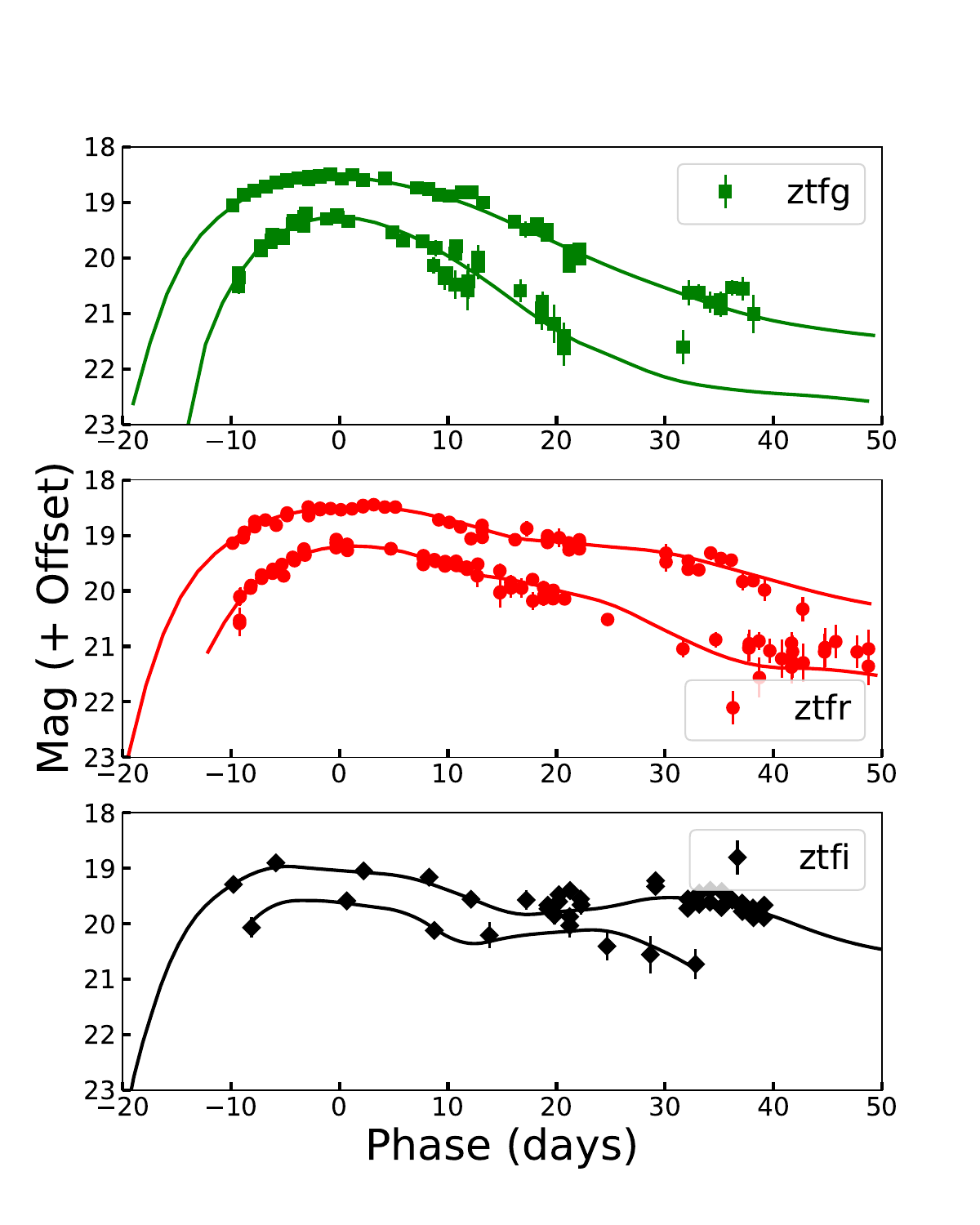}
    \vspace{-0.2cm}
    \caption{Lightcurves in the ZTF $g,r,i$ filters along with the SALT2 fits overplotted for the SN~Ia pair from the spec subsample with the largest difference in $x_1$, i.e. ZTF18abdmgab and ZTF20abqefja. As discussed in the text, the high $\Delta x_1$ (and low $\Delta c$) are important for constraining the width-luminosity relation. We can see the difference in $x_1$ in the lightcurve shapes of the two SNe~Ia, as well as the time of the second maximum in the $i$-band and the $r$-band shoulder.}
    \label{fig:lc_pair}
\end{figure}

\begin{equation}
    \mu = m_{\rm B} + \alpha x_1 - \beta c - M_{\rm B} - \delta_{\rm host},
\end{equation}

\noindent where the $\delta_{\rm host}$ term, if it to global properties of the host galaxy corresponds to the ``mass-step" \citep[e.g.][]{Scolnic2022}, and hence, ``falls-out" when inferring the standardisation relations since both SNe in each pair are in the same host galaxy. However, we note that several studies in the literature find a dependence of the SN~Ia luminosity of local environmental properties, which would not cancel out in our analysis \citep{Roman2018,Kelsey2021}.  We note that unlike previous studies which simultaneously marginalised over the SALT2 parameters and the standardisation relations, we fit them in successive steps, since no degeneracy was seen between $\alpha$, $\beta$ and the SALT2 parameters \citep[e.g., see][]{Biswas2022}.

We fit for the $\alpha$ and $\beta$ values, marginalising over the true $x_1$ and $c$ values.
If the observed stretch and colour differences are $\Delta x_1^o, \Delta c^o$, the true values can be written $\Delta x_1 = \Delta x_1^o + \delta x_1, \Delta c = \Delta c^o + \delta c$  where $\delta x_1$ and $\delta c$ are the deviations from the true differences. 
The distance modulus difference is given by,
\begin{align}
\Delta\mu &= \Delta m + \alpha\Delta x_1 - \beta\Delta c = 
\Delta m + \alpha\Delta x_1^o - \beta\Delta c^o + \alpha\delta x_1 - \beta\delta c\\
 &\equiv \Delta\mu^o + \alpha\delta x_1 - \beta\delta c.
\end{align}
Note that the uncertainty of $\Delta\mu^o$ is given $\sigma_{\Delta\mu^o} = \sigma_{\Delta m}$. This will include the measurement uncertainty in the observed magnitudes, the intrinsic magnitude dispersion and possible dispersions in $\alpha$ and $\beta$. The likelihood is now given by

\begin{multline}
L = \Pi \frac{1}{\sqrt{2\pi\sigma_m^2}}\exp{\left[-\frac{1}{2}\frac{(\Delta\mu^o + \alpha\delta x_1 - \beta\delta c)^2}{\sigma_m^2}\right]}\frac{1}{\sqrt{2\pi\sigma_{\delta x_1}^2}}\\
\exp{\left[-\frac{1}{2}
\frac{(\delta x_1)^2}{\sigma_{\delta x_1}^2}\right]}\frac{1}{\sqrt{2\pi\sigma_{\delta c}^2}}\exp{\left[-\frac{1}{2}\frac{(\delta c)^2}{\sigma_{\delta c}^2}\right]}.
\end{multline}
Substituting $k_1 = \alpha \delta x_1$ and $k_2 = \beta \delta c$ we get
\begin{multline}
L = \Pi \frac{1}{\sqrt{2\pi\sigma_m^2}}\exp{\left[-\frac{1}{2}\frac{(\Delta\mu^o + k_1 - k_2)^2}{\sigma_m^2}\right]}\frac{1}
{\sqrt{2\pi\alpha^2\sigma_{\delta x_1}^2}}\\
\exp{\left[-\frac{1}{2} 
\frac{k_1^2}{\alpha^2 \sigma_{\delta x_1}^2}\right]}\frac{1}{\sqrt{2\pi\beta^2\sigma_{\delta c}^2}}\exp{\left[-\frac{1}{2}\frac{k_2^2}{\beta^2\sigma_{\delta c}^2}\right]}
\end{multline}
the $\alpha$ and $\beta$ terms in the square root in the denominator are important to renormalise the likelihood correctly. Rewriting the $\sigma$ terms for brevity
\begin{align}
    a &\equiv \frac{(\Delta\mu^o)^2}{\sigma_m^2}\\
    b &\equiv \frac{\Delta\mu^o}{\sigma_m^2}\\
    c &\equiv \frac{1}{\sigma_m^2}\\
    c_\alpha &\equiv \frac{1}{\alpha^2\sigma_{\delta x_1}^2}\\
    c_\beta &\equiv \frac{1}{\beta^2\sigma_{\delta c^2}}
\end{align}
summing over all the pairs, gives the expression for the likelihood as  

\begin{multline}
L = \Pi \frac{1}{\sqrt{2\pi/c}}\frac{1}{\sqrt{2\pi/c_\alpha}}\frac{1}{\sqrt{2\pi/c_\beta}} \\ \exp{\left[-\frac{1}{2}\left(a + c(k_1-k_2)^2 + 2b(k_1-k_2) + c_\alpha k_1^2 + c_\beta k_2^2\right)\right]}
\end{multline}

Substituting $q_1 = k_1 + (b - k_2 c)/(c + c_\alpha)$ and then $q_2 = k_2 - b c_\alpha/(c c_\alpha + c c_\beta + c_\alpha c_\beta)$ gives us, 
\fontsize{7.8}{8}\selectfont
\begin{multline}
L = \Pi \frac{1}{\sqrt{2\pi/c}}\frac{1}{\sqrt{2\pi/c_\alpha}}\frac{1}{\sqrt{2\pi/c_\beta}} \\
\exp{\left[-\frac{1}{2}\left(a - \frac{b^2(c_\alpha + c_\beta)}{(c c_\alpha + c c_\beta + c_\alpha c_\beta)}   + q_1^2 (c + c_\alpha) + q_2^2\frac{(c c_\alpha + c c_\beta + c_\alpha c_\beta)}{(c + c_\alpha)}\right)\right]}
\end{multline}
\normalsize

Now, we integrate over $q_1$ and $q_2$ from minus to plus infinity, 

\begin{multline}
\int_{-\infty}^{\infty}\exp{-\frac{q_1^2 (c + c_\alpha)}{2}}dq_1 = \\
\sqrt{\frac{2\pi}{(c + c_\alpha)}} 
\int_{-\infty}^{\infty}\exp{-\frac{q_2^2 (c c\alpha + c c_\beta + c_\alpha c_\beta)}{2 (c + c_\alpha)}}dq_2 = \\
\sqrt{\frac{2\pi (c + c_\alpha)}{(c c_\alpha + c c_\beta + c_\alpha c_\beta)}}
\end{multline}

substituting into the expression for the likelihood gives us
\small
\begin{multline}
L = \Pi \sqrt{\frac{c c_\alpha c_\beta}{2\pi (c c_\alpha + c c_\beta + c_\alpha c_\beta)}}
\exp{\left[-\frac{1}{2}\left(a - \frac{b^2(c_\alpha + c_\beta)}{(c c_\alpha + c c_\beta + c_\alpha c_\beta)}\right)\right]},
\end{multline}
\normalsize
and 
\tiny
\begin{multline}
\chi^2 = -2\log (L) = \sum a - \frac{b^2(c_\alpha + c_\beta)}{(c c_\alpha + c c_\beta + c_\alpha c_\beta)} + \\ \log\left[\frac{2\pi (c c_\alpha + c c_\beta + c_\alpha c_\beta)}{c c_\alpha c_\beta}\right].
\end{multline}
\normalsize


\begin{figure}
    \centering
    \includegraphics[width=.5\textwidth]{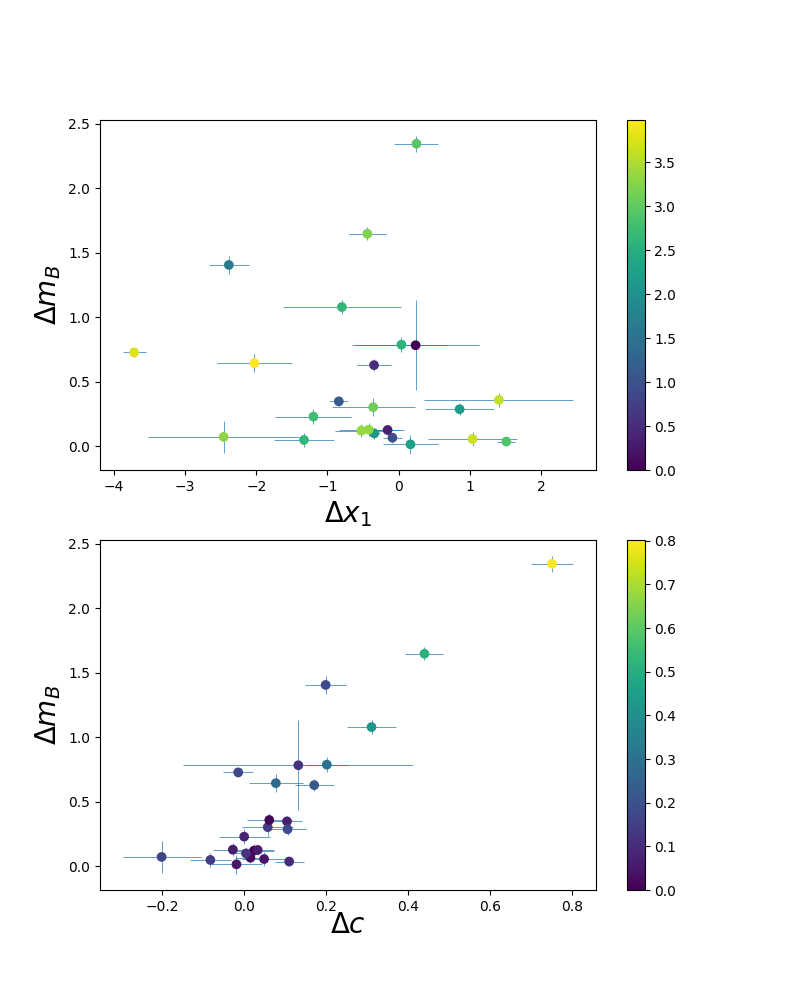}
    \caption{The difference in the inferred SALT2 $m_B$ versus the difference in the inferred $x_1$ (top) and $c$ (bottom) for each siblings pair in the sample. The siblings with large differences in $x_1$ (and similar values of $c$) predominantly constraint $\alpha$ precisely whereas those with large $\Delta c$ constrain $\beta$.  The color bar shows the $x_1$ for the wider SN (i.e. higher $x_1$; top) and $c$ for the redder (i.e. higher $c$; bottom) SN~Ia in the pair. }
    \label{fig:deltamb_x1c}
\end{figure}
\begin{table*}
\caption{SALT2 fit parameters for 
SN~Ia-SN~Ia pairs in this study both for the spectroscopic (spec) and photometric-spectroscopic (photo-spec) subsample.}
\fontsize{5.8}{8}\selectfont
\begin{tabular}{|c|c|c|c|c|c|c|c|c|c|c|c|}
\hline
SN1 & SN2 & Ang. Sep. (``) & z & $m_B$-1 &  $m_B$-2 & $x_1$-1    & $x_1$-2 & $c$-1 &  $c$-2 &  log (Mass) \\
\hline
&&&&& {\Large Spec} &&&& \\
&&&&  &&&&& \\
ZTF20abmarcv\_1 & ZTF20abmarcv\_2 & 0.0 & 0.1144 & 18.975 +/- 0.044 & 19.275 +/- 0.069 & 0.622 +/- 0.271 & 0.123 +/- 0.529 & -0.002 +/- 0.041 & 0.053 +/- 0.041 &9.44  \\
ZTF18abnucig & ZTF20achyvas & 1.7 & 0.09 & 18.802 +/- 0.043 & 18.682 +/- 0.038 & 0.339 +/- 0.209 & 0.887 +/- 0.304 & -0.075 +/- 0.036 & -0.099 +/- 0.036 &10.99 \\
ZTF22abveefy & ZTF21abnfdqg & 2.1 & 0.038 & 17.270 +/- 0.068 & 18.259 +/- 0.053 & -2.802 +/- 0.42 & -3.089 +/- 0.312 & -0.099 +/- 0.069 & 0.215 +/- 0.069 &10.63  \\
ZTF20acehyxd & ZTF21abouuow & 2.8 & 0.035 & 18.149 +/- 0.05 & 16.500 +/- 0.037 & 0.285 +/- 0.238 & 0.677 +/- 0.09 & 0.440 +/- 0.039 & -0.004 +/- 0.039 &10.03 \\
ZTF20aaeszsm & ZTF20abujoya & 2.9 & 0.07 & 18.564 +/- 0.039 & 18.614 +/- 0.048 & 0.136 +/- 0.312 & -1.206 +/- 0.274 & 0.066 +/- 0.035 & -0.012 +/- 0.035 &11.25 \\
ZTF20abptxls & ZTF21aabpszb & 3.3 & 0.0163 & 15.197 +/- 0.036 & 15.235 +/- 0.269 & 0.776 +/- 0.086 & -1.929 +/- 2.026 & 0.085 +/- 0.031 & -0.120 +/- 0.031 &9.82 \\
ZTF20aaxicpu & ZTF21abasxdp & 4.9 & 0.0721 & 18.647 +/- 0.033 & 18.583 +/- 0.034 & -1.575 +/- 0.096 & -1.478 +/- 0.098 & -0.052 +/- 0.029 & -0.068 +/- 0.029 &11.0 \\
ZTF19accobqx & ZTF19acnwelq & 8.7 & 0.09 & 18.584 +/- 0.039 & 18.529 +/- 0.049 & 1.19 +/- 0.386 & 0.159 +/- 0.462 & -0.032 +/- 0.035 & -0.075 +/- 0.035 &9.94 \\
ZTF20abatows & ZTF20abcawtk & 9.7 & 0.0945 & 19.113 +/- 0.033 & 19.074 +/- 0.033 & 0.42 +/- 0.093 & -1.087 +/- 0.089 & 0.022 +/- 0.026 & -0.089 +/- 0.026 &10.81 \\
ZTF20abydkrl & ZTF20acpmgdz & 30.9 & 0.0311 & 16.569 +/- 0.038 & 16.466 +/- 0.033 & -0.556 +/- 0.067 & -0.315 +/- 0.041 & 0.045 +/- 0.031 & 0.042 +/- 0.031 &11.37 \\
ZTF19abjpkdz & ZTF19aculypc & 45.4 & 0.0564 & 18.901 +/- 0.037 & 17.456 +/- 0.08 & -3.01 +/- 0.222 & -0.847 +/- 0.211 & 0.106 +/- 0.032 & -0.110 +/- 0.032 &11.56 \\
ZTF18abdmgab & ZTF20abqefja & 53.6 & 0.0802 & 19.386 +/- 0.034 & 18.658 +/- 0.033 & -2.225 +/- 0.14 & 1.312 +/- 0.106 & 0.085 +/- 0.029 & 0.099 +/- 0.029 &10.92 \\
&&&&  &&&& \\
&&&&& {\Large Photo-Spec} &&&& \\
&&&&  &&&&& \\
ZTF19acbzdvp\_1 & ZTF19acbzdvp\_2 & 0.0 & 0.103 & 19.401 +/- 0.039 & 19.114 +/- 0.041 & -0.309 +/- 0.207 & -1.086 +/- 0.446 & 0.098 +/- 0.032 & -0.008 +/- 0.032 &10.05 \\
ZTF19aambfxc\_1 & ZTF19aambfxc\_2 & 0.0 & 0.0541 & 19.880 +/- 0.062 & 17.531 +/- 0.035 & 0.424 +/- 0.273 & 0.214 +/- 0.130 & 0.734 +/- 0.046 & -0.026 +/- 0.046 &10.33 \\
ZTF19abaeyln & ZTF20abeadnl & 2.3 & 0.0852 & 18.582 +/- 0.041 & 19.484 +/- 0.051 & 0.390 +/- 0.284 & -1.719 +/- 0.314 & 0.113 +/- 0.034 & 0.108 +/- 0.034 &10.6 \\
ZTF20abazgfi & ZTF19acgemxh & 2.5 & 0.09 & 18.848 +/- 0.038 & 19.929 +/- 0.056 & 0.158 +/- 0.158 & -0.563 +/- 0.779 & 0.032 +/- 0.033 & 0.348 +/- 0.033 &10.36 \\
ZTF20abgaovd & ZTF19abtuhqa & 2.7 & 0.1 & 18.503 +/- 0.056 & 18.286 +/- 0.045 & -0.981 +/- 0.325 & 0.232 +/- 0.407 & -0.008 +/- 0.047 & 0.004 +/- 0.047 &10.66 \\
ZTF18aakaljn & ZTF19acdtmwh & 3.0 & 0.0699 & 18.35 +/- 0.051 & 18.996 +/- 0.066 & 1.553 +/- 0.391 & -0.455 +/- 0.398 & 0.122 +/- 0.047 & 0.207 +/- 0.047 &10.94 \\
ZTF22aaksdvi & ZTF21acowrme & 7.9 & 0.0821 & 18.306 +/- 0.034 & 18.674 +/- 0.063 & -0.306 +/- 0.129 & 1.138 +/- 0.964 & -0.127 +/- 0.029 & -0.057 +/- 0.029 &10.84  \\
ZTF18abuiknd & ZTF20acqpzbo & 8.6 & 0.104 & 19.085 +/- 0.036 & 18.96 +/- 0.047 & 0.401 +/- 0.281 & 0.854 +/- 0.330 & -0.044 +/- 0.033 & -0.016 +/- 0.033 &10.75 \\
ZTF20abgfvav & ZTF18abktzep & 9.1 & 0.095 & 19.116 +/- 0.037 & 19.903 +/- 0.057 & 0.027 +/- 0.191 & 0.048 +/- 0.609 & 0.025 +/- 0.032 & 0.227 +/- 0.032 &9.59 \\
ZTF19aatzlmw & ZTF20aaznsyq & 11.0 & 0.073 & 18.354 +/- 0.078 & 18.317 +/- 0.045 & -0.131 +/- 0.364 & -0.426 +/- 0.175 & -0.037 +/- 0.054 & -0.033 +/- 0.054 &10.73 \\
ZTF21aajfpwk & ZTF19aacxwfb & 17.8 & 0.0791 & 18.99 +/- 0.032 & 18.858 +/- 0.039 & -2.174 +/- 0.111 & -2.049 +/- 0.204 & -0.011 +/- 0.027 & -0.053 +/- 0.027 &11.38 \\
ZTF18aaqcozd & ZTF19aaloezs & 21.1 & 0.073 & 18.451 +/- 0.033 & 18.793 +/- 0.034 & -1.305 +/- 0.111 & -2.129 +/- 0.086 & -0.108 +/- 0.028 & -0.003 +/- 0.028 &10.87  \\
ZTF20abrgyhd & ZTF19aatvlbw & 30.1 & 0.066 & 18.447 +/- 0.033 & 19.067 +/- 0.045 & -1.791 +/- 0.107 & -2.224 +/- 0.222 & -0.024 +/- 0.029 & 0.148 +/- 0.029 &11.1 \\

\hline
    \end{tabular}
\label{tab:saltparams}
\end{table*}

Here, the $\sigma_{\rm fit}$ error term is derived from the output covariance matrix of the SALT2 model fit, for a given value of $\alpha$ and $\beta$. $\sigma_{\rm int}$ is the intrinsic scatter term and $\sigma_{\alpha, \beta}$ are the dispersions in the sample of $\alpha$ and $\beta$, respectively. $\Delta x_1$ and $\Delta c$ are the difference between the $x_1$ and $c$ for each sibling pair.

In our inference, we fit  $\alpha$, $\beta$, as well as their dispersions, as free parameters with uninformative priors.  For the default analysis we fit with five free parameters, including the intrinsic scatter. In alternate analysis cases, e.g. where the sample is fitted with a single $\alpha$ and $\beta$, we set the $\sigma_{\alpha, \beta} = 0$. We use \texttt{PyMultiNest} \citep{2014A&A...564A.125B}, a python wrapper to \texttt{MultiNest} \citep{2009MNRAS.398.1601F} to derive the posterior distribution on the parameters. We use the sampling efficiency optimal for parameter inference and 1200 live points.

\section{Results}
\label{sec:results}

In this section, we present SN~Ia lightcurve fit parameters and the inferred value of the luminosity-colour and luminosity-lightcurve width standardisation relations.
We fitted the SALT2 model to the lightcurves for our sample. To create the final sample for computing the standardisation relations, we remove all objects with an error on time of maximum $\sigma(t_0) > 2$\,days and with only observations in a single filter. Furthermore, we remove objects without adequate sampling at early phases. To quantify this selection criterion, we use the best sampled SNe, i.e., ZTF20acpmgdz, ZTF20achyvas to perform a test of  recovering the SALT2 parameters by downsampling the data and fitting in the absence of early time data. We find that for cases with data at least 3 days before maximum we can recover the $x_1$ and $c$ values from the full lightcurve, however, if the first observation is at a later epoch, the values are biased by $> 2 \sigma$  compared to the inference from the full lightcurve. We, therefore, only select pairs where both the SNe have at least one observation before -3 days, to avoid any biases in the $\alpha$ and $\beta$ measurements from biased $x_1$ and $c$ inference. This leaves us with 12 spec and 13 photo-spec SN~Ia pairs, a total of  25 pairs. 

The parameter distributions are shown in Figure~\ref{fig:param_dist} and reported in Table~\ref{tab:saltparams}. Since the aim is to constrain $\alpha$ and $\beta$ and not cosmological parameters, we do not make selection cuts on the value of $x_1$ and $c$, allowing for the full range of observed lightcurve widths and colours in our sample. For comparison, in Figure~\ref{fig:param_dist}, we plot the complete parameter distribution of the ZTF DR2 sample as dashed lines (S24, R24). While the siblings do not extended to the highest redshifts in the DR2 distribution, they span the observed range of $x_1$ and $c$ values of the entire DR2 sample. Note that for direct comparison we plot the DR2 sample without the cosmological cuts on $x_1$ and $c$. While the $c$ distribution has a high p-value (0.11) for a Kolmogorov-Smirnov (KS) test between the DR2 and sibling samples, the $x_1$ distribution has a low p-value (0.014) suggesting that even though the siblings span the entire range of observed $x_1$ values in the DR2 sample, the distribution is not drawn from the same parent population. We note from Figure~\ref{fig:param_dist} that the mass distribution for the siblings is skewed towards higher values compared to the values for the DR2 distribution. This would be expected for a siblings sample since it is more likely that larger galaxies produce two SNe~Ia. This may suggest that since more massive older galaxies typically host low $x_1$ events we see, on average, more siblings that have lower $x_1$ \citep{rigault2020,Nicolas2021}. 
\begin{figure*}
    \centering
    \includegraphics[width=.8\textwidth]{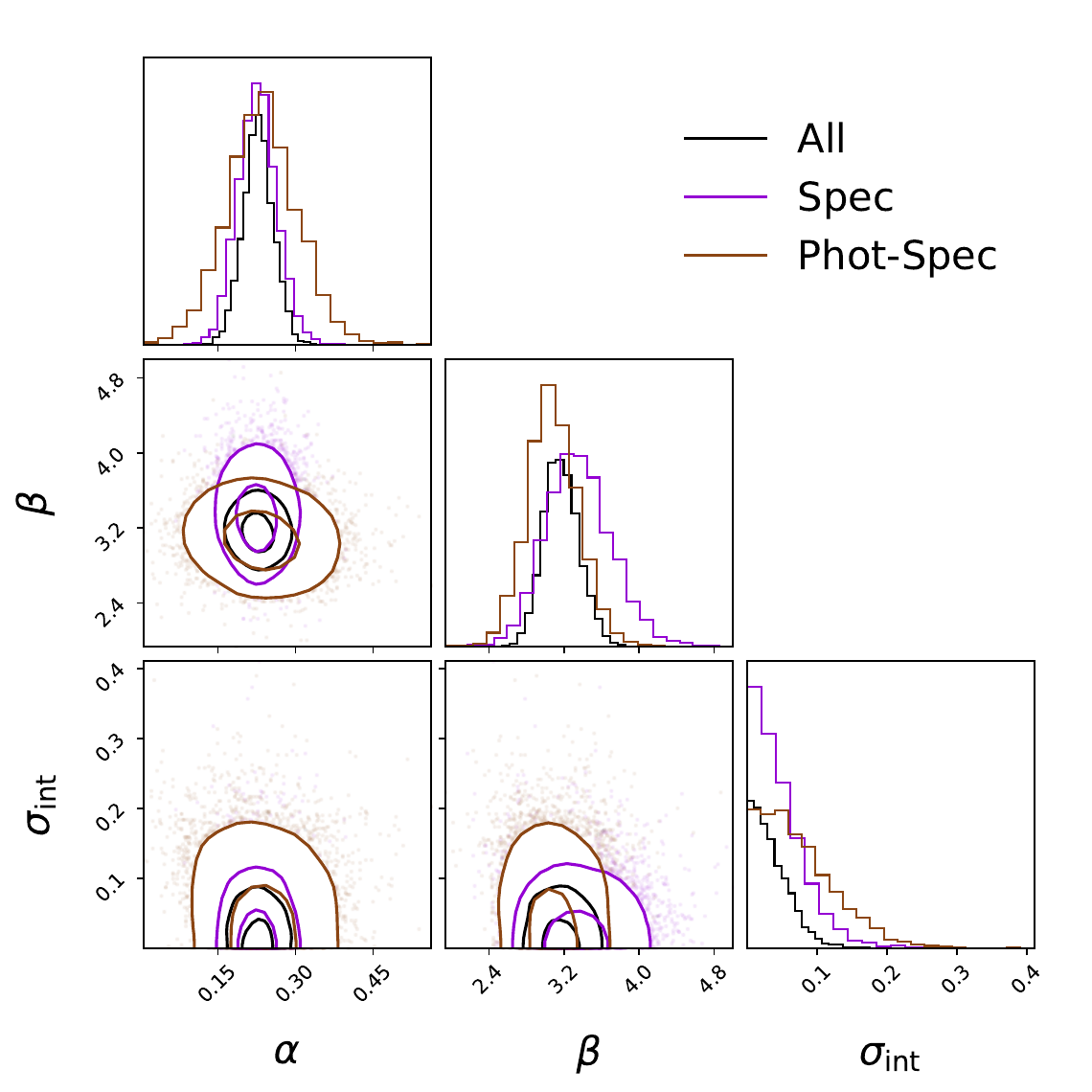}
    \caption{Constraints on $\alpha$, $\beta$ and the intrinsic dispersion ($\sigma_{\rm int}$) for the complete (black), spectroscopic (violet) and photometric (brown) samples.}
    \label{fig:alpha-beta-spread}
\end{figure*}

From Figure~\ref{fig:deltamb_x1c}, we find that the difference in the peak magnitude is correlated more significantly with the difference in colour than the lightcurve width (the pearson correlation coefficient $r=0.42$ compared to $r=-0.045$). Therefore, we expect stronger constraints on the $\beta$ parameter, however, for our fiducial fit, we simultaneously infer $\alpha$ and $\beta$. 

The sample has a wide distribution of angular separations for the siblings pairs. In both the spec-spec and phot-spec samples, there are three sibling pairs each where the separation is smaller than the pixel size of the camera, and hence, these are ``same pixel" siblings, similar to the pair presented in \citet{Biswas2022}.  This is interesting, since the small separation would also indicate that the difference in the properties of the local environment of the SN is small.

When fitting with the dispersion in $\alpha$ and $\beta$ (equation~\ref{eq:chisq}), we obtain $\alpha = 0.218  \pm   0.055$ and $\sigma(\alpha) \leq  0.195$ and on $\beta = 3.084 \pm 0.312$ and $\sigma(\beta)  \leq 0.923$, where $\sigma_{\alpha, \beta}$ are the dispersions in $\alpha$ and $\beta$. Below we evaluate constraints on the standardisation relations and their dispersion for the individual subsamples, i.e. spec and phot-spec respectively.

\begin{figure}
    \centering
    \includegraphics[width=0.48\textwidth]{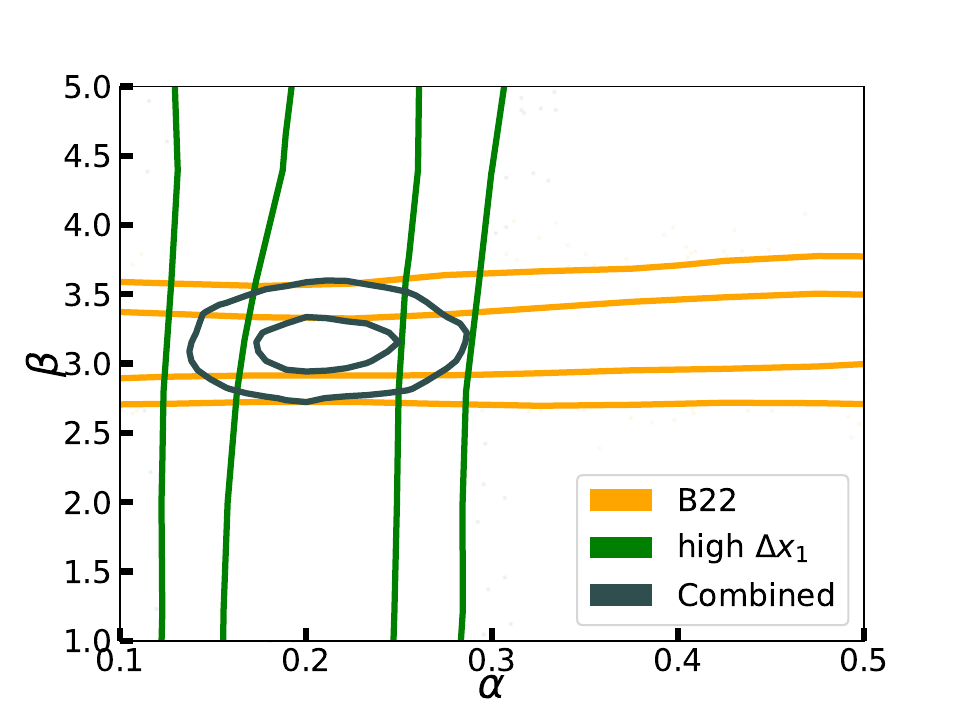}
    \caption{The constraints on $\alpha$ and $\beta$ from two sibling pairs. The $\beta$ constraints are from the pair analysed in \citet{Biswas2022}  (high $\Delta c$, low $\Delta x_1$) and the $\alpha$ constraints are from the pair ZTF18abdmgab-ZTF20abqefja (high $\Delta x_1$, low $\Delta c$). The figure illustrates the orthogonality in the constraints from high $\Delta c$ (low $\Delta x_1$) and high $\Delta x_1$ (low $\Delta c$) sibling pairs. The combined constraints in black are for illustrative purposes, the final constraints on $\alpha$ and $\beta$ from the combined sample are more stringent than presented here.}
    \label{fig:al-bet-twopairs}
\end{figure}
Combining all the pairs to constrain $\alpha$ and $\beta$ under the assumption of a single $\alpha$ and $\beta$, i.e. with the dispersion set to zero, we get $\alpha= 0.228  \pm  0.030$ and $3.162  \pm   0.191$ (Figure~\ref{fig:alpha-beta-spread}, black contours)

\subsection{Spectroscopic SN~Ia sample}
\label{ssec:spec_ia}
The SALT2 fit parameters for the spec sample are summarised in Table~\ref{tab:saltparams}. Unlike the cosmological sample, we do not make selection cuts on the measured value of $x_1$ and $c$. Therefore, the sample has a greater range of observed properties. The diversity of the entire SN~Ia sample from ZTF DR2 is discussed in a companion paper (Dimitriadis et al. in prep.). The range of $x_1$ and $c$ parameters  a long baseline to fit for the standardisation parameters. For one of the pairs with high $\Delta c$ (i.e. a difference $> 0.1$), ZTF20acehyxd+ZTF21abouuow, the colour excess attributed to extinction from Milky Way dust ($E(B-V)_{\rm MW}$) is also significant, i.e. 0.463 mag. We, therefore, test the assumption of the MW $R_V$ on the inferred $\beta$ constraint from this SN~Ia pair. We vary the MW $R_V$ to the line of sight from 2.5 to 3.5 and find no significant shift in the inferred $\beta$ value. Hence, for our analysis we continue to adopt the fiducial MW $R_V = 3.1$. 

Since our aim is to infer both the standardisation parameters and their dispersion in the sample, we fit the entire spec sample together with the likelihood expressed in equation~\ref{eq:chisq}. We find a value of $\alpha = 0.217  \pm  0.061$ and $\beta = 3.084  \pm  0.740 $.  This would suggest a mean $R_V \sim 2.2$ for the sample. Both the $\alpha$ and $\beta$ dispersion have a high median though it is consistent within 1$\sigma$ with no dispersion. 
  
We note that as expected, the constraints on $\alpha$ are driven by the sibling pairs that have a high $\Delta x_1$ (and similar $c$) and the constraints on $\beta$ by pairs that have high $\Delta c$ (and similar $x_1$). The similarity in the $c$ values while large differences in $x_1$ allow us to break the degeneracy between the $\alpha$ and $\beta$ constraints, since otherwise, if $\Delta x_1$ and $\Delta c$ were both large  there would be a strong correlation between inferred $\alpha$ and $\beta$.   




\subsection{Photometric-spectroscopic SN~Ia pairs subsample}
\label{ssec:photo_ia}
Along with the sibling pairs of spectroscopically confirmed SNe~Ia analysed in section~\ref{ssec:spec_ia}, ZTF has also discovered several pairs of SNe in the same galaxy where one SN in the pair is a spectroscopically confirmed SN~Ia and the other is a likely SN~Ia based on its lightcurve. 
We fit the photo-spec pairs with the same method as the spectroscopic pairs and report the SALT2 fit parameters values in Table~\ref{tab:saltparams}.
Similar to the spec sample in section~\ref{ssec:spec_ia}, infer $\alpha$, $\beta$ and the dispersion. We find $\alpha = 0.186  \pm  0.091$ and $\beta = 3.031  \pm  0.501$. Similar to the spectroscopic subsample, the median dispersion value is high, however, it is consistent with 0 at 1$\sigma$. The constraints are shown as brown contours in Figure~\ref{fig:alpha-beta-spread}.  We find that the parameter inference for the phot-spec sample is consistent with the spec sample and hence, we can combine the samples to get the most precise constraints on $\alpha$, $\beta$ and the dispersion.


\begin{figure}
    \centering
    \includegraphics[width=.48\textwidth]{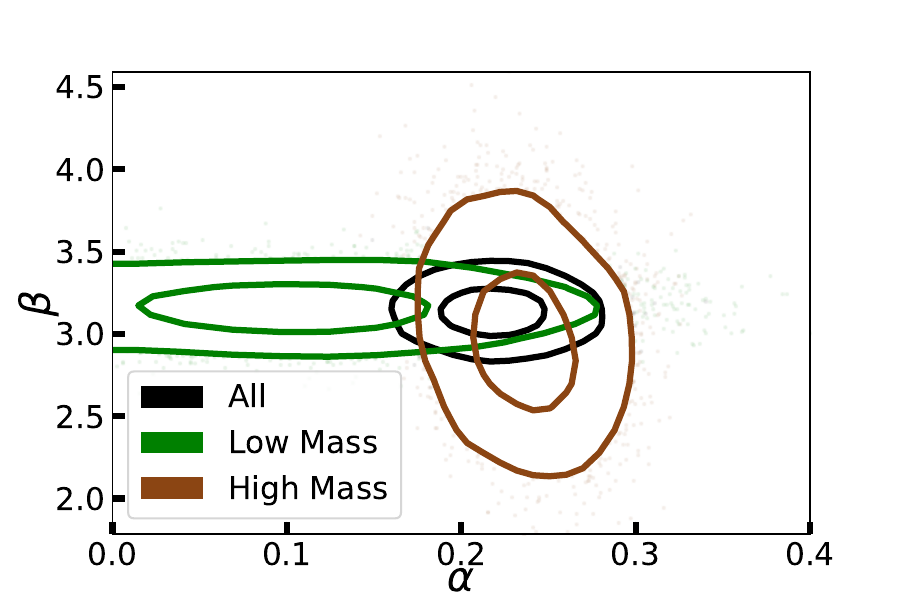}
    \caption{$\alpha$ and $\beta$ constraints from sibling pairs in low (green) and high (brown) mass host galaxies split at the median  ${\rm log}(M_{*}) = 10.57$. The values are consistent at 1.5 $\sigma$ between the subsamples. For comparison, the constraints from the entire sample are plotted as well (black).}
    \label{fig:mass_split}
\end{figure}   

\subsection{Host Galaxy Mass Dependence}
Recent studies have demonstrated that the reddening relations possibly depend on the properties of the host galaxy \citep[e.g.][]{BS20,Gonz_lez_Gait_n_2021}. 
Here, we test whether there is a dependence of the inferred $\alpha$ and $\beta$ and their dispersions on the host galaxy properties of the sibling pair. From the above results, we find that spec and photo-spec samples yield consistent values, hence, we combined both subsamples for this analysis to gain more statistical precision. 

We compute the stellar masses from the $g-i$ colour and $i$-band absolute magnitude of the host galaxy using the relation provided in \citet{Taylor2011} given as 
\begin{equation}
    {\rm log} \left(\frac{M}{M_\odot}\right) = 1.15 + 0.7 ({ m_g} - { m_i}) - 0.4  M_{i}
\label{eq:t11_host}
\end{equation}
 where $m_g$ and $m_i$ are MW extinction corrected apparent magnitudes in the $g$ and $i$ bands and $M_i$ is the absolute $i$-band magnitude.

To analyse the dependence, we split the sample into high and low mass pairs, similar to analyses in the literature with Hubble residuals \citep[e.g.][and references therein]{Brout2022,Johansson2021}. Dividing the sample into low and high mass bins at ${\rm log}(M_{*}/M_{\odot}) = 10$, we find consistent results between the two subsamples. However, the statistics in the low mass bin are significantly smaller than for the high mass bin. We, therefore, split at ${\rm log}(M_{*}/M_{\odot}) = 10.57$ and also find that the $\alpha$ and $\beta$ values are consistent at $1.5 \sigma$.

\begin{table*}
\centering 
\label{tab:alphabeta}
\caption{The mean $\alpha$ and $\beta$ along with the dispersion in both parameters for the complete, spectroscopic only and photo-spec sample. We report the values for the cases with a free $\sigma(\alpha)$ and $\sigma(\beta)$ as well as the cases with a fixed $\sigma(\alpha) = \sigma(\beta) = 0$. We also report the case with a split for $\alpha$ constraints based on the $x_1$ value, the two $\alpha$ values are for the low and high $x_1$ SNe~Ia respectively (see text for more details). } 
\begin{tabular}{|c|c|c|c|c|c|}
\hline
Sample & $\alpha$ & $\beta$ & $\sigma(\alpha)$ & $\sigma(\beta)$ & $\sigma_{\rm int}$ \\
\hline
All & 0.218  $\pm$  0.055 & 3.084  $\pm$  0.312 & $<$ 0.195 & $<$ 0.923 & $<$ 0.103 \\ 
Spec & 0.217  $\pm$  0.061 & 3.084  $\pm$  0.740 & $<$ 0.215 & $<$ 2.698 & $<$ 0.137 \\
Phot-Spec & 0.186  $\pm$  0.091 & 3.031  $\pm$  0.501 & $<$ 0.364 & $<$ 1.773 & $<$ 0.165 \\
&& Single $\alpha$, $\beta$ && \\
All & 0.228  $\pm$  0.030 & 3.162  $\pm$  0.191 & $\ldots$ & $\ldots$ & $<$ 0.088 \\
Spec & 0.226  $\pm$  0.038 & 3.345  $\pm$  0.351 & $\ldots$ & $\ldots$ & $<$ 0.119 \\
Phot-Spec & 0.235  $\pm$  0.069 & 3.075  $\pm$  0.277 & $\ldots$ & $\ldots$ & $<$ 0.177 \\
&& Split $x_1$ &&  \\
All & 0.274  $\pm$  0.045/0.134  $\pm$  0.072 & 3.164  $\pm$  0.187 & $\ldots$ & $\ldots$ & $<$ 0.097 \\
\hline
\end{tabular}
\end{table*}
\vspace{-0.2cm}

\section{Discussion}
\label{sec:discussion}
We have constrained the standardisation relations of SNe~Ia with both the spec and photo-spec subsamples and a joint analysis with all the sibling pairs.  Accurately constraining the standardisation relations is key to improving cosmological constraints with current and future SN~Ia datasets \citep{Brout2022,SRD_LSST}. 


One open question regarding the value of $\alpha$ and $\beta$ is whether they have a unique value for all SNe~Ia, or whether there is diversity in the values across the populations, specially whether it is correlated to, e.g. host galaxy properties \citep{BS20,Johansson2021,Wiseman2023}. 
We note that in the sample of SN~Ia siblings presented here, we can constrain the values of $\alpha = 0.228  \pm  0.030 $ and $\beta = 3.162  \pm  0.191$. The value of $\alpha$ is 2.3 $\sigma$ higher than the inference for the cosmological sample from the recent Dark Energy Survey results \citep[DES;][]{Vincenzi2024,DES_SN_2024} and $\sim 3 \sigma$ higher than the value from the Pantheon+ compilation \citep{Brout2022}. The value for $\beta$ is consistent ( $< 1 \sigma$ difference) with the inference from the cosmological analysis. We infer the diversity in both $\alpha$ and $\beta$ by inferring the values along with  $\sigma(\alpha)$ and $\sigma(\beta)$ term multiplying the $x_1$ and $c$ difference in the error term while fitting for the parameters. For the total sample, we infer a $\sigma(\alpha) \leq 0.195$ and $\sigma(\beta) \leq 0.923$, at the 95\% C.L.  

We note that the median value of $\alpha$ and $\beta$ for the fiducial case is consistent with the value from the fit to the cosmological SN~Ia sample \citep{Brout2022}. As a cosmology independent method, the SN~Ia siblings are a robust consistency check of the SN~Ia standardisation relations. From Table~\ref{tab:alphabeta}, we see that for the subsamples, while the central value of the dispersion can be high, it is still consistent with no dispersion in $\alpha$ and $\beta$ at the $\sim 1 \sigma$ level. For comparison, we also fit the individual SNe~Ia in the siblings pairs with a cosmology-dependent method (although without any cuts on $x_1$ and $c$ for the sample), i.e. from the minimising the scatter of the Hubble residuals, as is done for cosmological analyses. We find $\alpha= 0.21 \pm 0.03$ and $\beta = 2.84 \pm 0.28$ when fitting the Hubble residuals, which is consistent with the approach from fitting the sibling pairs in a cosmology independent way. Compared to previous analyses inferring $\alpha$ and $\beta$ in from siblings in a cosmology independent way \citep[e.g.][]{Biswas2022}, this analyses constrains both $\alpha$ and $\beta$ from the siblings alone, as opposed to only constraints on $\beta$ from previous work. Moreover, the constraint from $\beta$ has a 60$\%$ improvement in the uncertainty compared to previous studies, with a more conservative method to estimate uncertainties.

We, therefore, fit both the spec and phot-spec samples with only a single $\alpha$ and $\beta$ for the entire population. We note that both the individual subsamples have consistent $\beta$ values with the inference from the cosmological sample. However, we find a higher $\alpha$ value at the $\sim 2.5  \sigma$ level. 
We test whether there is evidence from the sibling sample, for a difference in $\alpha$ between subsamples based on $x_1$. We divide the sample based on the $x_1$ of each SN~Ia in a sibling pair to constrain $\alpha_{\rm low}$ and $\alpha_{\rm high}$, i.e. a single pair can be fitted with a different $\alpha$ if the $x_1$ values for the individual SNe are on different sides of the break.  We choose a break value of  $x_1 = -0.49$, based on the analysis of the standardisation of the entire ZTF second data release (DR2) sample of SNe~Ia in \textcolor{red}{Ginolin et al. in prep.} (G24). Fitting this broken power law, the low $x_1$ subsample has an $\alpha_{\rm low} = 0.274  \pm  0.045$ and $\alpha_{\rm high} = 0.133 \pm   0.072$. We perturb the break value ranging from -1 to 0, including the median value of -0.27, and do not find significant differences in the inferred $\alpha$ values. A detailed study of the standardisation effect is being conducted in a companion paper (G24).

We compare the $\alpha$ and $\beta$ values we get with the cosmological value from \citet{BS20}. We simulate a sample like the sibling pairs we observed from the intrinsic colour-luminosity relation ($\beta_{\rm int}$) and dust properties (e.g., total to selective absorption, $R_V$), inferred in \citet{BS20}, convolving the expected diversity in both $\beta_{\rm int}$ and $R_V$. Inferring a $\beta$ from the two effects combined, which is what we are fitting with the sibling sample, we find the value we get is consistent with the $\beta$ for the cosmological sample.  We perform a similar comparison for our sibling subsamples split by the host galaxy masses. For the low mass subsample we find consistency with the cosmological sample, however, the constraints have large errors as the sample size is small. For the high mass subsample, we find that the $\beta$ corresponds to a larger $R_V$ (taking $\beta \sim R_V + 1$) that, while the mean for the high mass subsample in \citet{BS20} by $\sim 3 \sigma$, when convolving with the dispersion in the cosmological sample, we find that the value from the siblings analysis is within range. 

We note that our conclusions are not affected by splitting the sample at close to the median mass value of ${\rm log}(M_{*}/M_{\odot}) = 10.57$ (Figure~\ref{fig:mass_split}). We note that in all the cases studied here, we find that the intrinsic scatter in the sibling sample, when doing a pairwise comparison is smaller than the typically observed scatter in the cosmological sample, with a median scatter of $\sigma_{\rm int} = 0.047$ and the 95 $\%$ C.L. upper limit of $\sigma_{\rm int} \leq 0.088$ mag. This has also been seen in the literature sample of 12 pairs analysed by \citet{Burns2020}. The recent study of \citet{Dwomoh2023} analysed SN~Ia siblings in the near infrared (NIR) and found evidence for residual intrinsic scatter, however, they suggest that it could possibly arise from a need for better observations and reduction in the NIR.
 
With future surveys, like the Vera C. Rubin Observatory's Legacy Survey of Space and Time (LSST), we expect to find $\sim 800$ SN~Ia siblings \citep{Scolnic2020}. Assuming a similar fraction of $\sim 10 - 20\%$ of the sibling pairs have high $\Delta x_1$ or high $\Delta c$, which is the driving factor for constraining $\alpha$ and $\beta$, we can expect a $\sim$ factor of 3 improvement in the uncertainty on $\alpha$ and $\beta$. Such a sample would also be crucial to confirm or refute the ``break" in $\alpha$ between low and high $x_1$ SNe~Ia (see also G24). The stated improvements would make the future siblings constraints comparable to the current best cosmological constraints \citep{Brout2022}. Given the rates of siblings from LSST, it is possible that a small subsample would also contain $> 2$ SNe~Ia, i.e. SN~Ia triplet \citep[e.g.][]{Ward2023}, which can be important, depending on the shape and colour of the SNe to constrain both $\alpha$ and $\beta$ from a single host galaxy. 
\vspace{-0.2cm}
\section{Conclusions}
\label{sec:conclusion}

In this study, we analysed a uniformly observed sample of sibling SNe~Ia, i.e. multiple SNe in the same parent galaxy, from the Zwicky Transient Facility. This is the single largest sample of SN~Ia sibling pairs, observed with a single instrument and photometric system, allowing us to reduce the uncertainties on the inferred standardisation parameters, $\alpha$ and $\beta$. Our sample contains a total of 25 pairs with 12 pairs having a spectroscopic classification for both SNe~Ia and 13 pairs where one object has been spectroscopically classified as an SN~Ia and the sibling is a photometricly classified SN~Ia. Interestingly, three of the 25 pairs (and a further 3 that didn't pass the quality cuts) were discovered with very small separations within the host-galaxy, and observed on the same pixel on the detector. 

We infer $\alpha$ and $\beta$ in a cosmology-independent way, by comparing the standardisation of both siblings in the pair. This method does not require a computation of a cosmological distance or a correction of the observed redshift for peculiar motions in the local universe since both these quantities are the same for each sibling in the pair. For the fiducial analysis, we infer both $\alpha, \beta$ and the spread in their values for the two subsamples. We find that the spec subsample indicates an $\alpha = 0.217  \pm  0.061$  and $\beta = 3.084  \pm  0.740$ and the photo-spec sample indicates $\alpha = 0.186  \pm  0.091$ and $\beta = 3.031  \pm  0.501$. Both subsamples point to a median $\beta$ value that is consistent with cosmological analysis, and points towards an $R_V \sim \beta - 1$ that is significantly lower than the canonical Milky Way value of 3.1. These results are consistent with the findings from a single sibling pair of \citet{Biswas2022}. However, we note that with a large permissible dispersion in the $\beta$ values it is likely that an individual galaxy can have consistent dust properties with that of the Milky Way.

While the fiducial analysis yield a high dispersion value for both $\alpha$ and $\beta$, the uncertainties on the dispersion  parameters are large enough that the samples could be consistent with having only a single $\alpha$ and $\beta$. We, therefore, constrain $\alpha$ and $\beta$ using only a single linear relation without any dispersion and find $\alpha = 0.228  \pm  0.030 $ and $\beta = 3.162 \pm 0.191$. 
We also subdivided the sample based on host galaxy mass, into the canonical low and high mass bins split at ${\rm log}(M_{*}/M_{\odot}) = 10.57$. The $\alpha$  and $\beta$ values for the subsamples are consistent, showing no strong host galaxy dependence.  

Future surveys like LSST are expected to find $\sim 800$ SN~Ia siblings, which will be an excellent sample to improve the uncertainties on $\alpha$ and $\beta$ and understand the diversity in the width-luminosity and colour-luminosity relations \citep{Scolnic2020}.

\section*{Acknowledgements}
SD acknowledges support from the Marie Curie Individual Fellowship under grant ID 890695 and a Junior Research Fellowship at Lucy Cavendish College. This work has been supported by the research project grant “Understanding the Dynamic Universe” funded by the Knut and Alice Wallenberg Foundation under Dnr KAW 2018.0067.
AG acknowledges support from  {\em Vetenskapsr\aa det}, the Swedish Research Council, project 2020-03444.
This project has received funding from the European Research Council (ERC) under the European Union's Horizon 2020 research and innovation programme (grant agreement n°759194 - USNAC)
L.G. acknowledges financial support from the Spanish Ministerio de Ciencia e Innovaci\'on (MCIN), the Agencia Estatal de Investigaci\'on (AEI) 10.13039/501100011033, and the European Social Fund (ESF) "Investing in your future" under the 2019 Ram\'on y Cajal program RYC2019-027683-I and the PID2020-115253GA-I00 HOSTFLOWS project, from Centro Superior de Investigaciones Cient\'ificas (CSIC) under the PIE project 20215AT016, and the program Unidad de Excelencia Mar\'ia de Maeztu CEX2020-001058-M, and from the Departament de Recerca i Universitats de la Generalitat de Catalunya through the 2021-SGR-01270 grant.
JHT and KM acknowledge support from EU H2020 ERC grant no. 758638.
Based on observations obtained with the Samuel Oschin Telescope 48-inch and the 60-inch Telescope at the Palomar Observatory as part of the Zwicky Transient Facility project. ZTF is supported by the National Science Foundation under Grants No. AST-1440341 and AST-2034437 and a collaboration including partners Caltech, IPAC, the Weizmann Institute of Science, the Oskar Klein Center at Stockholm University, the University of Maryland, Deutsches Elektronen-Synchrotron and Humboldt University, the TANGO Consortium of Taiwan, the University of Wisconsin at Milwaukee, Trinity College Dublin, Lawrence Livermore National Laboratories, IN2P3, University of Warwick, Ruhr University Bochum, Northwestern University and former partners the University of Washington, Los Alamos National Laboratories, and Lawrence Berkeley National Laboratories. Operations are conducted by COO, IPAC, and UW. SED Machine is based upon work supported by the National Science Foundation under Grant No. 1106171. The ZTF forced-photometry service was funded under the Heising-Simons Foundation grant \#12540303 (PI: Graham). The Gordon and Betty Moore Foundation, through both the Data-Driven Investigator Program and a dedicated grant, provided critical funding for SkyPortal.

\section*{Data Availability}
All data associated with this publication will be made available via \texttt{github}  as part of the ZTF second data release of Type Ia supernovae.



\bibliographystyle{aa}
\bibliography{siblings_aa} 


\vspace{-0.2cm}
\appendix
\section{Sample Selection}
\label{sec:sample_select}
In this section, we present the full list of siblings discovered in the ZTF data stream. 
\begin{table*}
\caption{List of the sibling pairs from the cross-match query. }
\resizebox{.7\textwidth}{!}{
\begin{tabular}{|l|l|c|c|r|r|}
\hline
SN1 & SN2 & Separation (``) & Redshift & Host R.A.& Host Dec. \\
\hline
&&& &&\\
&& Spec &&& \\
&&& &&\\
ZTF20abmarcv  & ZTF20abmarcv  & 0.7 & 0.1144 & 311.3210 & 7.1135 
\\ 
ZTF18abnucig  & ZTF20achyvas & 1.7 & 0.09 & 293.41165 & 39.39022 \\ 
ZTF22abveefy  & ZTF21abnfdqg & 2.1 & 0.0380 & 32.761750 & 36.481611 \\
ZTF20acehyxd  & ZTF21abouuow  &  2.8 & 0.0350 & 24.76783 & 75.32419 \\ 
ZTF20aaeszsm  & ZTF20abujoya  & 2.9 & 0.07 & 59.92458 & 26.58892 \\
ZTF20abptxls  & ZTF21aabpszb  & 3.3 & 0.016348 & 18.517083 & -1.742278 \\
ZTF20aaxicpu  & ZTF21abasxdp  & 4.9 & 0.072067 & 209.67597 & 43.12438 \\
ZTF19accobqx  & ZTF19acnwelq  & 8.7 & 0.09 & 335.05070 & 17.61122 \\
ZTF20abatows  & ZTF20abcawtk  & 9.7 & 0.094516 & 243.53431 & 30.04226 \\
ZTF20abydkrl  & ZTF20acpmgdz  & 30.9 & 0.031141 & 66.585625 & -10.098444 \\ 
ZTF19abjpkdz  & ZTF19aculypc  & 45.4 & 0.056436 & 38.9996 & 10.4405 \\
ZTF18abdmgab  & ZTF20abqefja  & 53.6 & 0.08024 & 250.91845 & 33.53957 \\
&&&&&\\
&&&&&\\
&& {\bf Rejected} &&& \\
&&&&&\\
&&&&&\\
ZTF20abzetdf  & ZTF20abzetdf  & 0. & 0.07 & 77.4320 & 3.8930 \\ 
ZTF19acykjad  & ZTF19acykjad  & 0. & 0.0630 & 351.44290 & 32.87435 \\
ZTF19aaugoig  & ZTF22abmzete  & 5.9 & 0.050852 & 175.50302 & 26.91578 \\
ZTF21acempzi  & ZTF22aalsabr  & 6.4 & 0.036969 & 325.772201 & -17.543818 \\ 
ZTF18adaadmh  & ZTF20aamujvi  & 8.2 & 0.0445 & 18.7834 & 1.5018  \\ 
ZTF18acckoil  & ZTF20abnngbz  & 8.5 & 0.032109 & 37.98036 & 3.12560 \\ 
ZTF20abbhyxu  & ZTF20acebweq  & 9.9 & 0.0319161 & 243.44182 & 22.91900 \\ 
ZTF21abybgjx  & ZTF22aankymj  & 10.1 & 0.044678 & 346.0400 & -6.4218 \\ 
ZTF19aaxeetj  & ZTF20abxbjai  & 10.5 & 0.05508536 & 197.98927 & 44.81551 \\ 
ZTF20aayqjpv  & ZTF21abcxswe  & 11.1 & 0.032056 & 221.02966 & 18.01263 \\ 
ZTF22abanxam  & ZTF20aasctts  & 11.9 & 0.04566 & 241.25274 & 27.25826 \\
ZTF20acwfftd & ZTF21aavqphe & 25.5 & 0.02147 & 224.43705 & 6.62692 \\ 
\hline
&&& &&\\
&& Phot-Spec &&& \\
&&& &&\\
ZTF19acbzdvp$_1$  & ZTF19acbzdvp$_2$  & 0 & 0.103 & 18.410434 & 40.852562 \\ 
ZTF19aambfxc$_1$ & ZTF19aambfxc$_2$ & 0 & 0.0541 & 265.42960 & 67.96197 \\ 
ZTF19abaeyln  & ZTF20abeadnl  & 2.3 & 0.085243 & 231.27763 & 11.58622 \\ 
ZTF20abazgfi  & ZTF19acgemxh  & 2.5 & 0.081 & 278.0636 & 43.7956 \\ 
ZTF20abgaovd  & ZTF19abtuhqa  & 2.7 & 0.077 & 251.191667 & -1.324667 \\
ZTF18aakaljn  & ZTF19acdtmwh  & 3.0 & 0.069910 & 145.29339 & 24.02284 \\
ZTF22aaksdvi  & ZTF21acowrme  & 7.9 & 0.082067 & 304.496833 & -6.299972 \\
ZTF18abuiknd  & ZTF20acqpzbo  & 8.6 & 0.104 & 29.36162 & 8.91661 \\ 
ZTF20abgfvav  & ZTF18abktzep  & 9.1 & 0.086159 & 226.92461 & 32.00931 \\
ZTF19aatzlmw  & ZTF20aaznsyq  & 11.0 & 0.073 & 258.4956 & 3.4938 \\ %
ZTF21aajfpwk  & ZTF19aacxwfb  & 17.8 & 0.079139 & 151.89978 & 58.21146 \\
ZTF18aaqcozd  & ZTF19aaloezs  & 21.1 & 0.073212 & 190.55964 & 42.26644 \\
ZTF20abrgyhd  & ZTF19aatvlbw  & 30.1 & 0.066 & 317.52829 & 8.05601 \\
&&&&& \\
&& {\bf Rejected} &&& \\
&&&&& \\
&&&&& \\
ZTF20aamibse$_1$  & ZTF20aamibse$_2$  & 0 & 0.097034 & 214.82711 & 0.05774 \\ 
ZTF20aagnbpw  & ZTF20aaunioz  & 3.8 & 0.052359 & 203.10639 & 38.36005 \\ 
ZTF20aaznlnj  & ZTF19aaklsto  & 1.5 & 0.078 & 251.12719 & 52.81174  \\ %
ZTF19aazcxwk  & ZTF18abaidds  & 2.5 & 0.12 & 259.35255 & 45.43164 \\ 
ZTF21aaxvrva  & ZTF21abhqoja  & 3.2 & 0.081652 & 235.97511 & 26.24748  \\ 
ZTF18abzpbpi  & ZTF21aaletht  & 4.9 & 0.08930 & 135.13114 & 36.46106 \\ %
ZTF19aaksrgj  & ZTF20aavpwxl  & 5.5 & 0.0859496 & 189.52611 & 8.04589 \\ 
ZTF20abcgjvq  & ZTF19aaxpbdh  & 5.2 & 0.05608 & 310.93079 & -1.23889 \\ 
 %

ZTF21aagkvqa  & ZTF20abhttyd & 9.6 & 0.06 & 12.27975 & 18.25831 \\ 
ZTF19aakluwr  & ZTF20acpqbue  & 11.1 & 0.057997 & 42.247199 & 26.510890  \\ %
ZTF20abasewu  & ZTF20acdccnl  & 43.3 & 0.054923 & 12.39237 & 23.57823 \\ %
ZTF20acmgkqe  & ZTF17aadlxmv  & 2.2 & 0.061960 & 127.44817 & 33.90647 \\ 
ZTF18aazcoob  & ZTF19aalbqxs  & 9.7 & 0.084498 & 269.5105 & 69.0740 \\ 

ZTF22aaksdvi  & ZTF21acowrme  & 7.9 & 0.082067 & 304.496833 & -6.299972 \\ 
ZTF18aawmvbj  & ZTF21aagaehc  & 2.4 & 0.1403 & 153.17225 & 21.41557 \\ %
\hline
\end{tabular}}
\label{tab:sample}
\end{table*}


\end{document}